# Effective Absorption Enhancement in Small Molecule Organic Solar Cells by Employing Trapezoid Gratings *[1]

Xiang Chun-Ping (相春平)[a], Liu Jie-Tao (刘杰涛)[a], Jin Yu(金玉)[b], Xu Bin-Zong (许斌宗)[a], Wang Wei-Min (汪卫敏)[a], Wei Xin (韦欣)[a], Song Guo-Feng (宋国峰)[a], and Xu Yun (徐云)[a]**[2]

[a]Nano-optoelectronics Laboratory, Institute of Semiconductors, Chinese Academy of Sciences, A35, Tsinghua East Road, Beijing 100083, People's Republic of China

[b]State Key Laboratory on Integrated Optoelectronics, College of Electronic Science and Engineering, Jilin University, 2699 Qianjin Street, Changchun, 130012, People's Republic of China.

We demonstrate the optical absorption has been enhanced in the small molecule organic solar cells by employing trapezoid grating structure. The enhanced absorption is mainly attributed to both waveguide modes and surface plasmon modes, which has been simulated by using finite-difference time-domain method. The simulated results show that the surface plasmon along the semitransparent metallic Ag anode is excited by introducing the periodical trapezoid gratings, which induce high intensity field increment in the donor layer. Meanwhile, the waveguide modes result a high intensity field in acceptor layer. The increment of field improves the absorption of organic solar cells, significantly, which has been demonstrated by simulating the electrical properties. The simulated results exhibiting 31 % increment of the short-circuit current has been achieved in the optimized device, which is supported by the experimental measurement. The power conversion efficiency of the grating sample obtained in experiment exhibits an enhancement of 7.7 %.

Email: xuyun@semi.ac.cn, xiangcp@semi.ac.cn
Phone number: +86-010-82304164

---

*Supported by the National Nature Science Foundation of China under Grant Nos 61036010 and 61177070, along with the National Basic Research Program of China under Grant Nos 2011CBA00608 and 2012CB619203, and the National Key Research Program of China under Grant No 2011ZX01015-001.

**Corresponding author. Email: xuyun@semi.ac.cn

# 1. Introduction

Organic solar cells (OSCs) are of great interest in recent years for their unique properties of low-cost processing, mechanical flexibility and scalability, though their power conversion efficiency are lower than the traditional silicon-based solar cells.[1,2] Up to now, the reported efficiency of OSCs is up to 9.2 % by using tandem structure with different kinds of polymer materials,[3] and the efficiency of small molecule organic solar cells (SMOSCs) is up to 12 %.[4] However, the primary bottleneck of OSCs is the trade-off between exciton diffusion and light absorption.[5,6] The optical absorption length of organic materials is around 100 nm, while the exciton diffusion length is on the order of 10 -- 50 nm.[7-10] Therefore, the absorption efficiency and exciton diffusion efficiency in conventional OSCs are coupled, opposing quantities; increasing absorption by increasing the device thickness leads to a decrease in exciton diffusion efficiency and vice versa.[11] Additionally, due to the mismatch of the work function between organic materials and metal electrodes, the charge collection efficiency of OSCs is poor, which further limits the efficiency.[12] Hence, how to resolve these problems is premised.

Recently, many approaches have been proposed to release this trade-off of OSCs. Among, applying metallic subwavelength structure in OSCs is a promising way to trap the incident light into organic layers.[6,8,13-20] The main effect of the subwavelength structure is that it can support the surface plasmon polaritons (SPPs) propagating along the metal/dielectric interface and the localized surface plasmon (LSP).[5,7,19,21,22] For small metallic nanoparticles with dimension much smaller than the wavelength, the scattering efficiency is lower than the absorption efficiency, which results in obvious thermal loss.[13,23] In contrast to metallic nanoparticles, periodical metallic nanograting with a tunable period can modulate the surface plasmon resonances and scatter photons into the active layer.[8] In previous work, metallic nanograting had been used in traditional silicon-based solar cells to improve the light harvest efficiency, which has been demonstrated by the theoretical analysis.[11,17] In this paper, we propose a process-compatible trapezoid nanograting structure in top-illuminated SMOSCs to increase the optical absorption efficiency. We optimize the grating structure by calculating the absorption spectra of active layer, and reveal the reason for the absorption enhancement in different cases by analyzing the physical mechanisms. The simulated current density-voltage properties of the optimized device exhibit a significant improvement, which are further supported by experiment.

# 2. Model, calculation and experiment

## 2.1 Model and optical property

We investigate the effect of trapezoid nanograting in the top-illuminated SMOSCs. The model is calculated by using two-dimensional finite-difference time-domain (FDTD) method under the irradiation of transverse magnetic (TM) and transverse electric (TE) polarized light, separately. In our simulation, the grid size is set to 2 nm×2 nm and the perfectly matched layer and the periodic boundary condition are used in $z$ and $x$ direction, respectively. The schematic diagram of SMOSCs is shown in Fig. 1. The silver (Ag) with nanograting structure is adopted as the substrate. The biphasic calcium phosphate (BCP) is introduced as the exciton blocking layer on top of the cathode, and the fullerene ($C_{60}$) and subphthalocyanine (Subpc) are used as accepter and donor, respectively. A 20 nm-thick Ag anode is deposited on the top of the Subpc layer and presents a semitransparent property. The complex refractive indices of Subpc and BCP used for simulation are measured by spectroscopic ellipsometer and those of $C_{60}$ and Ag are employed from Ref. 24 and 25.

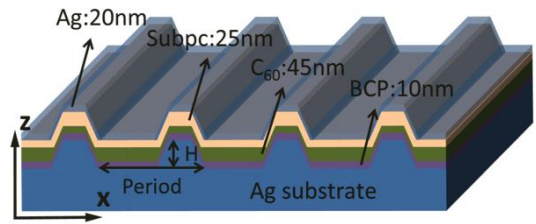

**Fig. 1** Schematic diagram of the trapezoid grating top-illuminated SMOSCs.

We first investigate the optical absorption properties of the proposed model. In order to characterize the absorption efficiency of SMOSCs, we particularly calculate the absorption of the active layer instead of the whole device. Fig. 2(a) shows the absorption coefficient of Subpc and $C_{60}$, which is defined as $\alpha=2\kappa(\omega)\omega/c$.[23] The $\kappa$ is the extinction coefficient, $\omega$ is the angular frequency of the incidence and $c$ denotes the velocity of light in free space. Two separated intrinsic absorption regions corresponding to the donor and accepter give a broadband absorption spectrum of active layer, which, on the other hand, need to be synthesized to get a maximum enhancement. We set the duty ratio of the grating as 0.5 and 0.33 at the bottom and top side of the trapezoid respectively, while the grating height and period are considered to be optimized. Since the grating height is a main parameter which influences the scattering effect,[26] we first simulate the unpolarized optical absorption spectra of organic layers with grating heights from 0 nm to 125 nm (Fig. 2(b)), where the period is set to be 300 nm. The absorption spectrum is enhanced significantly from 350 nm to 620nm, which is corresponding to the intrinsic absorption of Subpc and $C_{60}$. The spectrum reaches a maximum increment when the grating height reaches 100 nm, moreover, a larger grating height may decrease the optical absorption and electrical properties of the device.[17] To further understand the reason for the enhancement, the TE and TM polarized optical absorption spectra are calculated.

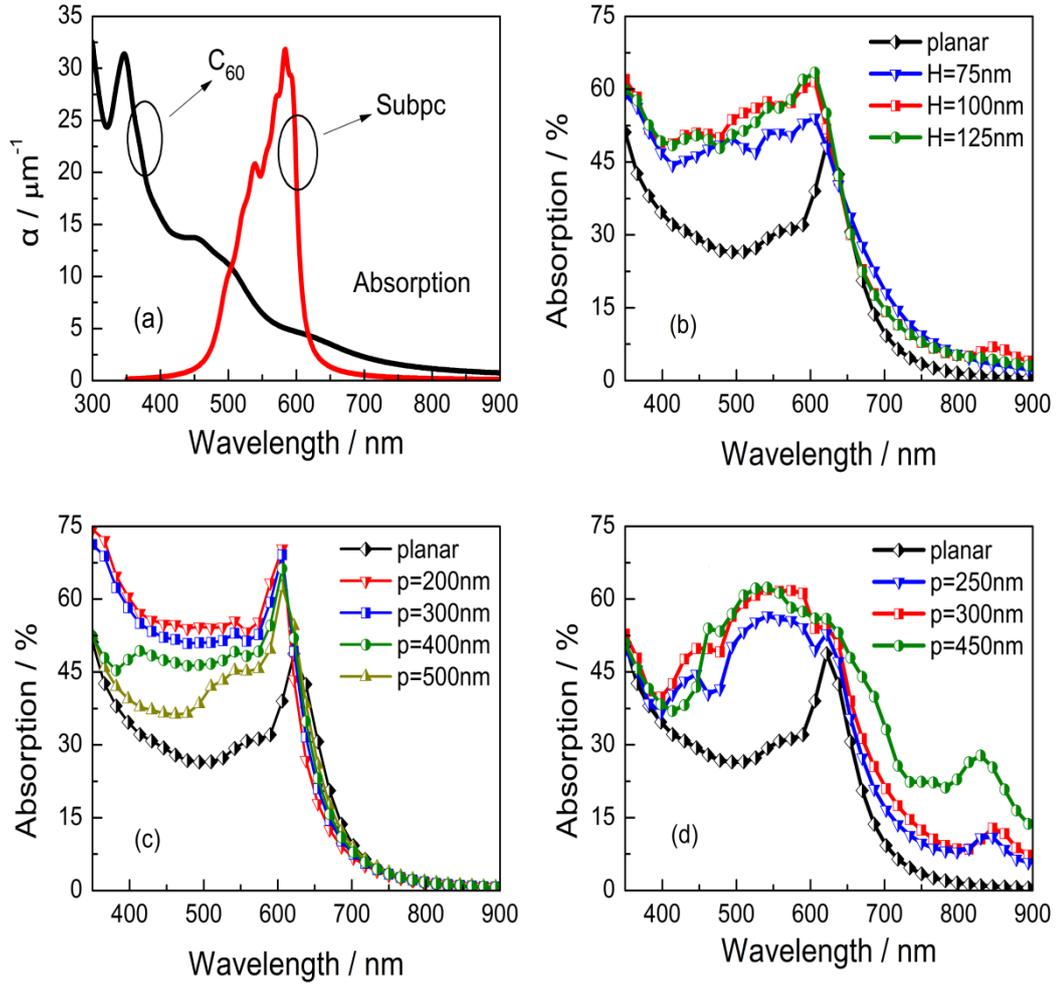

**Fig. 2** (a) The absorption coefficient of $C_{60}$ and Subpc. (b) Unpolarized optical absorption spectra of the sample with different grating heights. (c) TE polarized absorption spectra of the sample with different grating periods. (d) Absorption spectra of the sample with different grating periods for TM polarization.

As shown in Fig. 2(c), the TE polarized optical absorption efficiency of OSCs with trapezoid nanograting structure exhibits above 50 % at wavelength below 600 nm when the grating period is around 300 nm, which is higher than that of the planar OSCs with the same volume of active layer. The main reason for the TE polarized light absorption enhancement is the excitation of waveguide modes. By scattering the incident light into the active layer, the grating structure fulfills the phase matching condition of waveguide modes and increases the interaction of light with organic materials.[17] For normal incidence, the diffraction angle of the grating structure could be expressed as:

$$n_{org} \Lambda \cos(\psi) = m\lambda \qquad (1)$$

where the $n_{org}$ is the refractive index of active layer, $\Lambda$ is the period of grating, $\psi$ is the diffraction angle, $m$ denotes the diffraction order and $\lambda$ is the free space wavelength. The phase matching condition of the waveguide modes resonance is:

$$2k_{org}h - \varphi_1 - \varphi_2 = 2M\pi \qquad (2)$$

where the $k_{org}=2n_{org}\pi sin(\psi)/\lambda$ is the normal component of the propagating wave vector, $h$ is the thickness of waveguide layer, $\varphi_1$ and $\varphi_2$ are two polarization-depending phase shifts introduced by total internal reflection at the top and bottom surfaces of waveguide, $M$ denotes the mode order of waveguide. As indicated from equation 1 and 2, the resonant wavelength red shifts with the increased period. To further elucidate the waveguide modes in organic layers, the distribution of electric field are shown in Fig. 3. Due to the refractive indices of $C_{60}$ are higher than those of Subpc at wavelength below 550 nm, the electric field is mainly localized in $C_{60}$ layer at the resonant wavelength (Fig. 3(a)). Moreover, when the wavelength of incident light is smaller than the resonant wavelength, the electric field is hard to couple in the waveguide layer (Fig. 3(b)). Consequently, an absorption valley presents at the region smaller than the resonance wavelength, and extends toward long wavelength with the increased period (Fig. 2(c)). For the phase shifts $\varphi_1$ and $\varphi_2$ of waveguide modes in TM polarization are larger than those in TE polarization,[26] the valley in TM polarized absorption spectra presents at shorter wavelength (Fig. 2(d)).

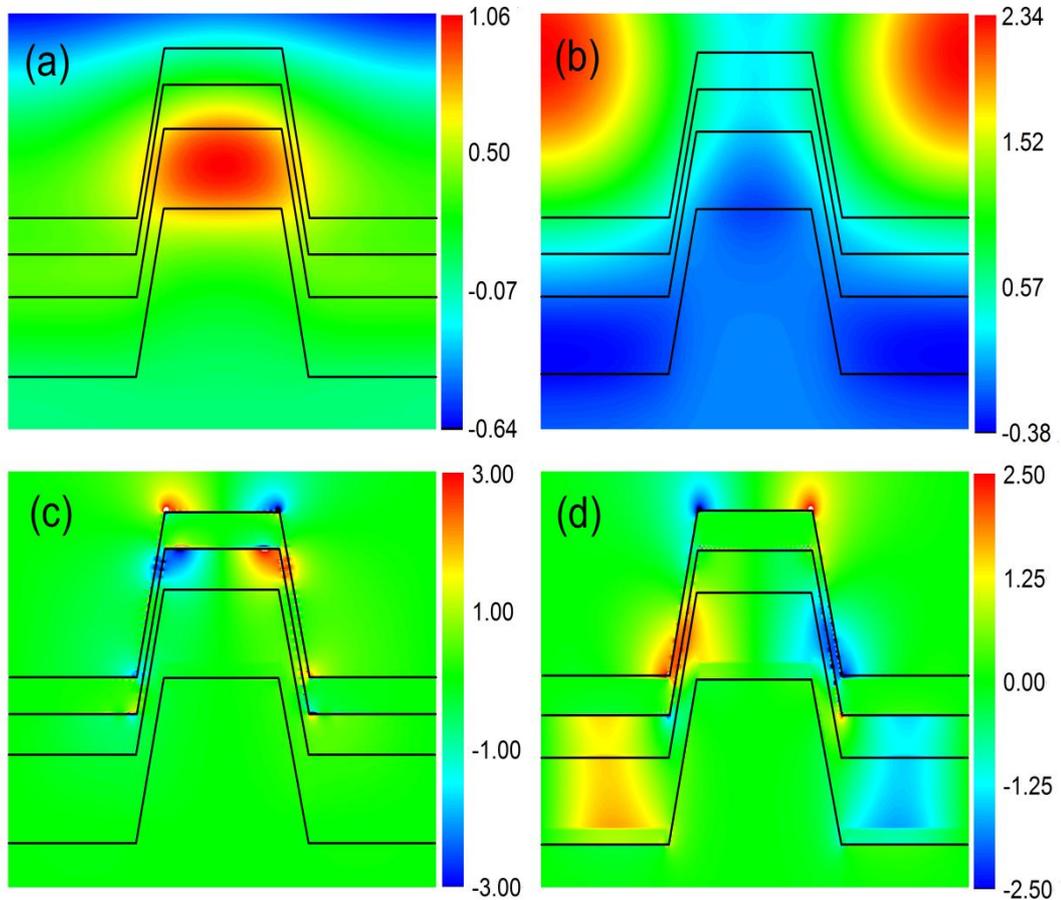

**Fig. 3** Distribution of electric field at wavelength of 400 nm with the period $\Lambda=250$ nm (a) and at the wavelength of 420 nm with the period $\Lambda=450$ nm (b) under TE polarized light, respectively. (c) and (d) correspond to the distribution of electric field at wavelength of 455 nm with the period $\Lambda=300$ nm and at wavelength of 830 nm with the period $\Lambda=450$ nm under TM polarized light, respectively.

Not only the waveguide modes, the SPPs and LSP can also be excited by the TM polarized light and increase the TM polarized absorption.[23] In the case of small period (under 300 nm), the LSP is the main reason for TM polarized absorption enhancement.[7] The LSP resonance can be excited directly by the incident light and the interaction of LSP at each unit cell comes into effect when the period decreases.[23] As the grating vector is perpendicular to the wave vector, the longitudinal LSP mode can be excited by the normal incident light around λ=450 nm and performs a red-shift with the increased period. The electric field localizes close to the anode at the top of the trapezoid grating at the longitudinal LSP resonance, which can increase the absorption of Subpc obviously (Fig. 3(c)). Due to the non-uniform dimension of the trapezoid with a narrow top and wide bottom in each unit cell, a broadband absorption enhancement corresponding to the absorption of Subpc (at wavelength 500 nm -- 600 nm) exhibits an absorption efficiency of above 60 %, which is twice higher than that of planar device. This increment arises from the excitation of LSP modes with the energy mostly localized on the bevel of the trapezoid, which could signally increase the current density since the active layer of the resonant region is thin for exciton diffusion.[17] As the grating period increases, the SPPs come into effect at long wavelength,[7] with the electric field mostly confined in the bottom of the grating (Fig. 3(d)). The absorption of the device with period $\Lambda$=450 nm exhibits an absorption enhancement at λ=830 nm comparing to that of the device which the period is smaller.

## 2.2 Current density-voltage character of the proposed device

To demonstrate the electrical performance of the proposed thin-film SMOSCs devices, we calculate the current density-voltage curves of the device under unpolarized AM 1.5G illumination. In our simulation, we assume that each photon generates one electric–hole pair. As the band gap of $C_{60}$ is about 1.9 eV[24] and the intrinsic absorption region of Subpc is from 450 nm to 650 nm, the photon with energy lower than 1.9 eV is hard to excite the exciton and transforms into heat ultimately. Thus we particularly consider the absorption spectrum between 350 nm and 700 nm. As is shown in Fig. 4, the short-circuit current density of the device with grating period $\Lambda$=300 nm increases to 7.37 mA/cm$^2$, 31 % higher than that of the planar device (5.62 mA/cm$^2$). The device with period $\Lambda$=450 nm shows a relatively lower current density since the waveguide modes cannot couple into the $C_{60}$ layer in the intrinsic absorption region.

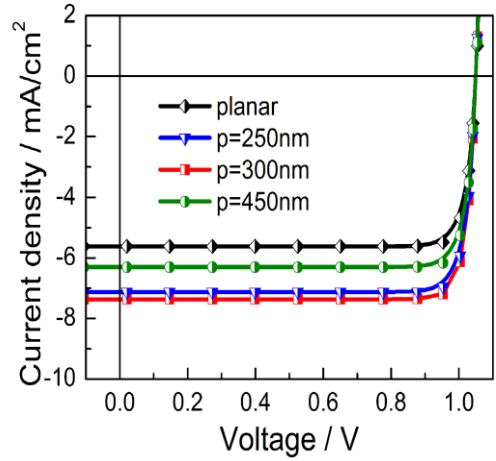

**Fig. 4** Simulated current density-voltage curves of the system for varied periods.

In order to confirm the simulation results, we prepare devices with an area of 2 mm×2 mm on the grating substrate. We firstly fabricate one-dimensional nanograting by using the interference lithography. A laser beam at λ=325 nm is used to generate the interference pattern on a 100 nm-thick photoresist layer, followed by the developing. The substrate is etched by the inductively coupled plasma (ICP) etching technology to form trapezoid gratings, and the scanning electron microscope (SEM) image of the grating structure is shown in Fig. 5(a). Afterward Ag (100 nm)/Ca (1 nm)/BCP (10 nm)/$C_{60}$ (45 nm)/Subpc (25 nm)/Ag (20 nm) are fabricated by vacuum thermal evaporation process sequentially and the thickness of each layer is monitored by quartz crystal film thickness monitor. The ultrathin Ca modify layer is deposited to form an ohmic contact, which can decrease the contact resistance without influencing the optical properties of the device. In order to demonstrate the enhancement effect, the OSCs device with planar structure is also fabricated as the reference sample. The current density-voltage character of the device are measured under 100 mW/$cm^2$ AM 1.5G illumination (Fig. 5(b)). By increasing the optical absorption (or alternatively, the generation rate of exciton) of the device, the short-circuit current density of the sample with grating structure increases to 3.93 mA/$cm^2$, 16.4 % higher than that of the planar device which is 3.38 mA/$cm^2$. The power conversion efficiency exhibits an enhancement of 7.7 % (increases from 1.51 % to 1.63 %). Meanwhile, the filling factor (FF) of the grating sample (39.8 %) shows a negligible decrement comparing with the planar one (41.4 %), indicating an acceptable processing compatibility of the trapezoid grating SMOSCs.

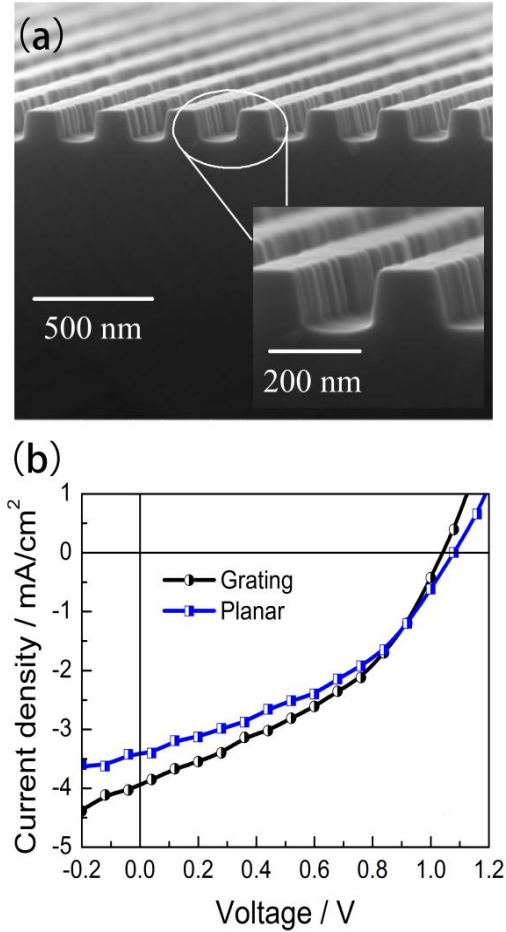

**Fig. 5** (a) SEM cross-sectional view of the grating. The grating height is 110 nm and the period is 300 nm. (b) Measured current density-voltage curves for the device with a grating period of 300 nm.

Comparing with the experiment, the photovoltaic properties of organic materials and the mismatch of the work function of different materials (i.e., the internal quantum efficiency and the charge collection efficiency) are not taken into account in our simulation, which makes the simulated current density higher than experiment result. Moreover, the larger interface area of grating sample increases the surface recombination hence leads to a lower charge collection efficiency comparing with that of the planar sample, which results in a decreased increment of short-circuit current density (from 31 % to 16.4 %).[17,27] The FF of devices, major difference between the simulation (about 88 %) and experiment (about 40 %), is mainly caused by the unavoidable leakage current and series resistance. The Ca modify layer can decrease the series resistance and suppress the decrement of FF. Nevertheless, the reduction in open-circuit voltage of the grating samples (Fig. 5(b)) indicates that the leakage current, which increases with the interface area and decreases the shunt resistance, is the main reason for the decrement of FF.[27] It is noted that understanding of the physical mechanisms could open up a new way to engineer and optimize other structures by this method. Moreover, a two-dimensional grating structure can be processed to realize a further improved light-trapping effect.

# 3. Conclusion

In conclusion, we investigate the effect of trapezoid grating on both TE and TM polarized optical absorption of top-illuminated SMOSCs. A significant enhancement caused by waveguide and LSP modes below 600 nm is observed, and the SPPs get excited around 800 nm. We improve the intrinsic absorption of the active layer by adjusting the grating period and grating height. The simulated and experimental short-circuit current density of the device with grating structure presents 31 % and 16.4 % improvement comparing to the planar device, respectively. The usage of the trapezoid grating in top-illuminated SMOSCs improves the power conversion efficiency by 7.7 % in experiment. The proposed design, simulation and fabrication can potentially be extended to different types of low-cost ultra-thin OSCs.